# Néel Spin Currents in Antiferromagnets

Ding-Fu Shao,[1,*,†] Yuan-Yuan Jiang,[1,2,*] Jun Ding,[3,*] Shu-Hui Zhang,[4] Zi-An Wang,[1,2] Rui-Chun Xiao,[5] Gautam Gurung,[6] W. J. Lu,[1] Y. P. Sun,[7,1,8] and Evgeny Y. Tsymbal[9,‡]

[1] *Key Laboratory of Materials Physics, Institute of Solid State Physics, HFIPS, Chinese Academy of Sciences, Hefei 230031, China*
[2] *University of Science and Technology of China, Hefei 230026, China*
[3] *College of Science, Henan University of Engineering, Zhengzhou 451191, People's Republic of China*
[4] *College of Mathematics and Physics, Beijing University of Chemical Technology, Beijing 100029, People's Republic of China*
[5] *Institute of Physical Science and Information Technology, Anhui University, Hefei 230601, China*
[6] *Trinity College, University of Oxford, OX1 3BH, United Kingdom*
[7] *High Magnetic Field Laboratory, HFIPS, Chinese Academy of Sciences, Hefei 230031, China*
[8] *Collaborative Innovation Center of Microstructures, Nanjing University, Nanjing 210093, China*
[9] *Department of Physics and Astronomy & Nebraska Center for Materials and Nanoscience, University of Nebraska, Lincoln, Nebraska 68588-0299, USA*

Ferromagnets are known to support spin-polarized currents that control various spin-dependent transport phenomena useful for spintronics. On the contrary, fully compensated antiferromagnets are expected to support only globally spin-neutral currents. Here, we demonstrate that these globally spin-neutral currents can represent the Néel spin currents, i.e. staggered spin currents flowing through different magnetic sublattices. The Néel spin currents emerge in antiferromagnets with strong intra-sublattice coupling (hopping) and drive the spin-dependent transport phenomena such as tunneling magnetoresistance (TMR) and spin-transfer torque (STT) in antiferromagnetic tunnel junctions (AFMTJs). Using $RuO_2$ and $Fe_4GeTe_2$ as representative antiferromagnets, we predict that the Néel spin currents with a strong staggered spin-polarization produce a sizable field-like STT capable of the deterministic switching of the Néel vector in the associated AFMTJs. Our work uncovers the previously unexplored potential of fully compensated antiferromagnets and paves a new route to realize the efficient writing and reading of information for antiferromagnetic spintronics.

Over decades, ferromagnetic metals have been intensively explored and widely employed in spintronics [1]. It has been demonstrated that their magnetization can be efficiently controlled by spin-torques [2] and used for generation of spin-polarized currents [1]. In particular, employing tunneling magnetoresistance (TMR) [3-5] and spin-transfer torque (STT) effects [6,7] in magnetic tunnel junctions (MTJs) allows electrical writing and reading of information that is stored in the magnetization orientation [8].

Recently, antiferromagnets have emerged as an alternative of ferromagnets for spintronic applications [9-12]. Due to being robust against magnetic perturbations, not producing stray fields, and exhibiting ultrafast spin dynamics, antiferromagnets may potentially overperform ferromagnets. However, reading and writing information using an antiferromagnetic (AFM) Néel vector as a state variable appears to be more challenging. Due to the absence of a net magnetization, spin-polarized currents and hence TMR and STT are not expected to occur in AFM systems.

In the past decade, significant progress has been achieved in the understanding of electronic and transport properties of antiferromagnets [13-20], resulting in discoveries of the spin-orbit torques [21-24], the linear [25-29] and nonlinear [30-34] anomalous Hall effects, the Néel vector dependent spin currents [35-47], and TMR in AFM tunnel junctions (AFMTJs) [48-50]. It was shown that AFM metals supporting nonrelativistic longitudinal spin-polarized currents driven by strong exchange interactions, rather than relatively weak spin-orbit coupling (SOC), can serve as an efficient spin source similar to ferromagnets or heavy metals, to exert a damping-like STT on antiferromagnets [35, 36, 51-56]. This damping-like STT can generate a persistent ultrafast oscillation under an applied driving current, which is promising for THz applications [51].

It is not obvious however how to use the advantages of such a uniformly polarized spin current to deterministically switch the Néel vector of an antiferromagnet. Such switching typically requires a uniform field-like spin-torque powered by a staggered spin-polarization on AFM sublattices [12, 21-23, 54], where the driving current is globally spin-neutral rather than uniformly spin-polarized. So far, such Néel spin-torques have occurred due to a relatively weak SOC and required devices with multiple in-plane terminals and thus large dimensions [10, 24]. It would be desirable to use the advantages of much stronger (compared to SOC) exchange interactions in AFM metals to generate a Néel STT capable of switching the Néel vector in the current-out-of-plane AFMTJs.

Recently, it was predicted that globally spin-neutral currents can generate prominent spin-dependent transport phenomena in AFMTJs, such as the TMR [48] and tunneling anomalous Hall effects [57]. These phenomena rely on the nonrelativistic exchange interactions in the bulk of AFM metals and their strengths are comparable to these in ferromagnetic systems. Importantly, these effects are robust against disorder and interface roughness, contrary to the previous predictions of STT and TMR in AFMTJs [58-63] hindered by disorder [64, 65]. One can expect therefore a possibility of a robust STT in AFMTJs driven by the AFM exchange interactions and globally spin-neutral currents.

In this paper, we predict the emergence of a Néel spin current, i.e. a staggered spin current flowing through different magnetic sublattices, in collinear fully compensated antiferromagnets. Based on symmetry analyses in real space, we find that the Néel spin currents are widely supported in AFM metals. Using first-principles quantum transport calculations [66], we predict strong Néel spin currents in $RuO_2$, an antiferromagnet with a spin-split Fermi surface, and $Fe_4GeTe_2$, an antiferromagnet with a spin-degenerate Fermi surface, and demonstrate a sizable field-like STT capable of the deterministic switching of the Néel vector in AFMTJs.

We consider a collinear AFM metal that consists of two magnetic sublattices $m_\alpha$, where $\alpha$ is the sublattice index, A or B (Fig. 1). A bias voltage ($V_b$) applied to this antiferromagnet generates a charge current $J$ composed of currents $J_{\alpha\beta}$ flowing through intra-sublattice ($\alpha = \beta$) and inter-sublattice ($\alpha \neq \beta$) so that

$$J_{\alpha\beta} = J_{\alpha\beta}^\uparrow + J_{\alpha\beta}^\downarrow, \qquad (1)$$

where ↑ and ↓ denote the two spin channels. The associated spin current $J_{\alpha\beta}^s$ is

$$J_{\alpha\beta}^s = J_{\alpha\beta}^\uparrow - J_{\alpha\beta}^\downarrow. \qquad (2)$$

The related conductance $g_{\alpha\beta}$ and spin conductance $g_{\alpha\beta}^s$ are

$$g_{\alpha\beta}^{(s)} = J_{\alpha\beta}^{(s)}/V_b. \qquad (3)$$

The $g_{\alpha\beta}^s$ is controlled by a symmetry operation $\hat{O}$ that connects the two magnetic sublattices. For example, many fully compensated antiferromagnets have combined $\hat{P}\hat{T}$ symmetry, where inversion symmetry $\hat{P}$ exchanges the two sublattices and time reversal symmetry $\hat{T}$ flips the magnetic moments. Applying $\hat{P}\hat{T}$ transformation to Eq. (3), we obtain $\hat{P}\hat{T}g_{AA}^s = -g_{BB}^s$ and $\hat{P}\hat{T}g_{AB}^s = -g_{AB}^s = 0$. Therefore, in $\hat{P}\hat{T}$ symmetric antiferromagnets, despite globally spin-neutral currents, staggered intra-sublattice spin currents are allowed (due to $J_{AA}^s = -J_{BB}^s$), while inter-sublattice spin currents are not (due to $J_{AB}^s = 0$). This observation is also valid for antiferromagnets compensated by combined $\hat{T}\hat{t}$ symmetry, where $\hat{t}$ is half a unit cell translation.

$\hat{O}$ can also be crystal symmetry, such as a mirror (glide) reflection or a rotation (screw rotation). For example, in a transport direction parallel to a mirror plane $\hat{M}$, $J_{\alpha\beta}^{(s)}$ represents a longitudinal $J_{\alpha\beta\parallel}^{(s)}$ current parallel to $\hat{M}$ or a transverse $J_{\alpha\beta\perp}^{(s)}$ current perpendicular to $\hat{M}$, with the related spin conductance components $g_{\alpha\beta\parallel}^{(s)}$ and $g_{\alpha\beta\perp}^{(s)}$. Similarly, we find that $\hat{M}$ enforces $\hat{M}g_{AA\parallel}^s = -g_{BB\parallel}^s$, $\hat{M}g_{AB\parallel}^s = -g_{AB\parallel}^s = 0$, $\hat{M}g_{AA\perp}^s = g_{BB\perp}^s$, and $\hat{M}g_{AB\perp}^s = g_{AB\perp}^s$. These relations imply the presence of a globally spin-neutral longitudinal current $J_\parallel$ that involves an intra-sublattice staggered longitudinal spin current $J_{AA\parallel}^s = -J_{BB\parallel}^s$ and a global transverse spin current $J_\perp^s = \sum_{\alpha,\beta} J_{\alpha\beta\perp}^s$. This symmetry analysis in real space is consistent with that previously performed in $k$-space [36, 48]. In particular, it explains the presence of the transverse spin current in $RuO_2$ for an electric field applied along [100], where there is only one mirror plane parallel to this direction [36, 40-42], and its absence for an electric field applied along [001], where there are multiple mirror planes parallel to this direction [48, 57]. Likewise, a global longitudinal spin current $J_\parallel^s$ is expected, if it does not have the symmetry constraints of $\hat{O}$ [36].

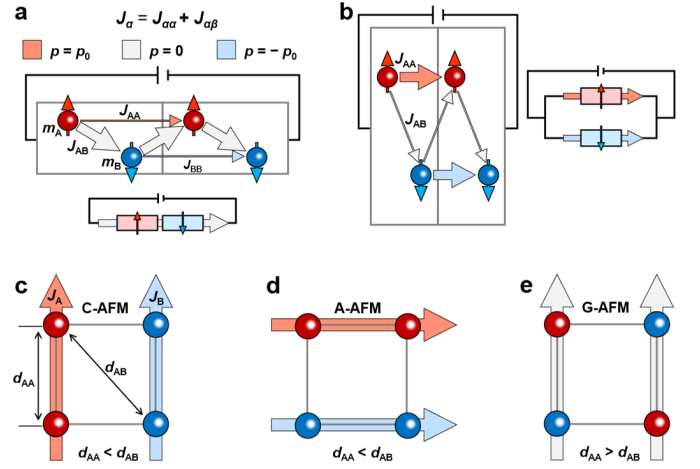

FIG. 1: (a) Schematics of a collinear AFM metal composed of sublattices A and B, where intra-sublattice distance $d_{AA}$ is larger than inter-sublattice distance $d_{AB}$. In this case, the intra-sublattice coupling is weak, and the antiferromagnet can be considered as a resistor with the two sublattices connected in series, where no spin current is flowing through each sublattice. (b) The same as (a) except $d_{AA} < d_{AB}$. In this case, the inter-sublattice coupling is weak, and the antiferromagnet can be considered as a resistor with the two sublattices connected in parallel, supporting a staggered Néel spin current. (c-e) Schematics of the Néel spin currents in C-type (c), A-type (d), and G-type (e) antiferromagnets.

The longitudinal Néel spin current through the magnetic sublattice $\alpha$ is given by

$$J_{\alpha\parallel}^s = \sum_\beta J_{\alpha\beta\parallel}^s. \qquad (4)$$

This current is, in general, spin-polarized with the spin polarization $p_{\alpha\parallel}$ defined by

$$p_{\alpha\parallel} = \frac{\sum_\beta J^s_{\alpha\beta\parallel}}{\sum_\beta J_{\alpha\beta\parallel}} = \frac{\sum_\beta g^s_{\alpha\beta\parallel}}{\sum_\beta g_{\alpha\beta\parallel}}. \quad (5)$$

When $g^s_{AB\parallel} = 0$, the magnitude of $p_{\alpha\parallel}$ is determined by the relative values of $g_{AB\parallel}$ and $g_{AA\parallel}$ ($g_{BB\parallel}$). Qualitatively, these conductances are determined by the inter- and intra-sublattice couplings (hoppings) that are related to the nearest inter- and intra-sublattice distances along the transport direction. When the inter-sublattice distance ($d_{AB}$) is much smaller than the intra-sublattice distance ($d_{AA}$) (Fig. 1(a)), the inter-sublattice coupling dominates, leading to a large $g_{AB\parallel}$ and hence small $p_{\alpha\parallel}$. In this case, the two AFM sublattices can be considered as being connected in a series with no spin current flowing through each sublattice. On the other hand, if $d_{AA}$ is much smaller than $d_{AB}$ (Fig. 1(b)), a small $g_{AB\parallel}$ and hence large $p_{\alpha\parallel}$ are expected. In this case, the two AFM sublattices are connected parallel supporting a Néel spin current.

Figures 1(c-e) illustrate typical AFM configurations for simple cubic systems. For C-type antiferromagnets composed of antiparallel-aligned ferromagnetic chains (Fig. 1(c)) and A-type antiferromagnets composed of antiparallel-aligned ferromagnetic layers (Fig. 1(d)), $d_{AA} < d_{AB}$ when the transport direction is parallel to the chains (layers), and thus Néel spin currents with sizable $p_{\alpha\parallel}$ are expected in this direction. On the other hand, the check-board-like magnetic order of G-type antiferromagnets resulting in $d_{AA} > d_{AB}$ (Fig. 1(e)) is not supportive to Néel spin currents.

To demonstrate a possibility of a strong STT, we consider the recently discovered high temperature antiferromagnet $RuO_2$ [67]. In its rutile structure, the ferromagnetically ordered $Ru_A$ and $Ru_B$ chains along the [001] direction are aligned antiparallel with a large interchain distance (Fig. 2(a)). Its magnetic space group $P4_2'/mnm'$ does not include $\hat{P}\hat{T}$ and $\hat{T}\hat{t}$ symmetries, but has two glide planes $\hat{M}_x$ and $\hat{M}_y$ perpendicular to the [100] and [010] directions, respectively, connecting the two sublattices $Ru_A$ and $Ru_B$. This results in the spin-split electronic structure of $RuO_2$ [68], generating sizable in-plane spin currents and associated spin-torques [36, 40-42], a giant TMR [48], and an anomalous Hall effect (if its Néel vector is rotated away from the easy axis) [29].

For $V_b$ applied along the [001] direction, a Néel spin current emerges in $RuO_2$, due to both $\hat{M}_x$ and $\hat{M}_y$ being parallel to [001]. $d_{AB}/d_{AA} = 1.14$ indicates a stronger intra-sublattice coupling, resulting in the Néel spin current with estimated $p_{\alpha\parallel} = 35\%$ [66, 69] that is comparable to the transport spin polarization of conventional ferromagnetic metals such as Fe, Co, and Ni [70-72]. This spin polarization is responsible for the predicted TMR in an AFMTJ based on $RuO_2$ electrodes [48].

The sizable $p_{\alpha\parallel}$ in $RuO_2$ also promises a strong STT [6,7]. To demonstrate this, we build an AFMTJ with $RuO_2$ (001) electrodes and a $TiO_2$ (001) insulating barrier layer. Figure 2(b) shows the atomic structure of the $RuO_2/TiO_2/RuO_2$ (001) supercell, which is used in our quantum-transport calculations and includes 6 $RuO_2$ layers on each side of the AFMTJ separated by 6 $TiO_2$ layers. To calculate STT exerted on the magnetic moments in the right electrode by the current from the left electrode, we assume that the Néel vectors are along the [100] and [001] directions in the left and right electrodes, respectively [73]. Figure 2(c,f) shows the calculated layer-resolved field-like STT $\tau_{FL,i}$ and damping-like STT $\tau_{DL,i}$ in the right half of the junction for a small $V_b = 0.1$ V. We find that $\tau_{FL,i}$ and $\tau_{DL,i}$ are large at the interfacial $RuO_2$ layers and then decrease rapidly at the layers away from the interface. This is consistent with the known behavior of STTs in MTJs [74]. Interestingly, $\tau_{FL,i}$ and $\tau_{DL,i}$ exhibit distinct behaviors: while $\tau_{FL,i}$ for the first two interfacial layers are parallel, $\tau_{DL,i}$ are antiparallel (staggered), indicating that the STTs are generated by the staggered spin polarization $p_{A\parallel} = -p_{B\parallel}$ of the Néel spin current. As a result, for the low $V_b$, the total $\tau_{FL}$ is large, while the total $\tau_{DL}$ is small (Fig. 2(e,h)).

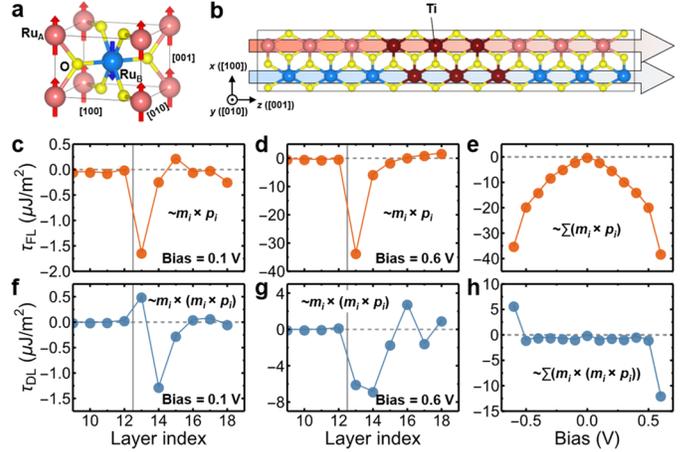

FIG. 2: **(a)** The atomic structure of $RuO_2$. **(b)** The atomic structure of the $RuO_2/TiO_2/RuO_2$ supercell. **(c-e)** The layer resolved field-like STT for $V_b = 0.1$ V (c) and $V_b = 0.6$ V (d) and the total field-like STT (e) in the right half of the $RuO_2/TiO_2/RuO_2$ AFMTJ. **(f-h)** The same as (c-e) for the damping-like STT. The vertical gray lines in (c,d,f,g) denote the interface between $TiO_2$ and $RuO_2$.

For a large $V_b = 0.6$ V, both $\tau_{FL,i}$ and $\tau_{DL,i}$ are enhanced. In particular, we find that $\tau_{DL,i}$ at the two interfacial layers become parallel. This is due to an additional uniform spin current with spin polarization $p_U$ generated by the uncompensated interfacial moments. Nevertheless, the total $\tau_{DL}$ remains small due to the most $\tau_{DL,i}$ being staggered. On the contrary, the total $\tau_{FL}$ is strongly enhanced and becomes comparable to that calculated for an Fe/MgO/Fe MTJ with similar barrier thickness [75]. Importantly, we find that independent of the interfacial configurations, $\tau_{FL}$ remains much stronger than $\tau_{DL}$, indicating the dominant role of the bulk Néel spin current for STT [66]. This fact implies that the STT should be robust against interface roughness.

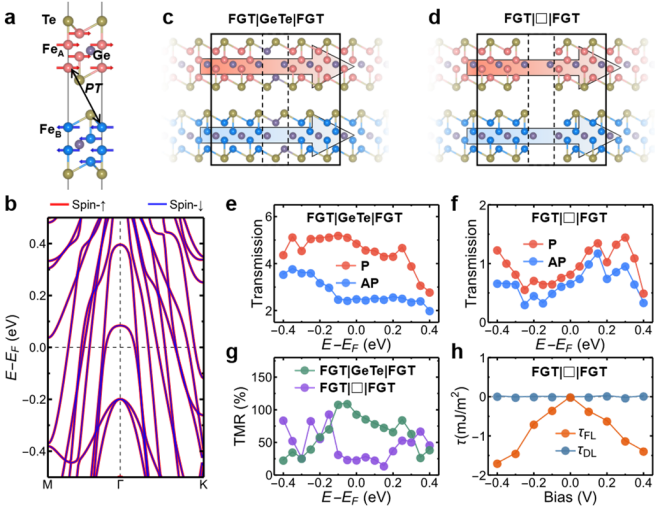

**FIG. 3:** **(a,b)** The atomic structure (a) and band structure (b) of an Fe$_4$GeTe$_2$ bilayer. **(c,d)** The supercells used for FGT|GeTe|FGT (c) and FGT|☐|FGT (d) AFMTJs. **(e,f)** The transmissions of FGT|GeTe|FGT (e) and FGT|☐|FGT (f) AFMTJs for the P and AP states as a function of energy. **(g)** The TMR ratio for FGT|GeTe|FGT and FGT|☐|FGT AFMTJs as a function of energy. **(h)** The total field-like and damping-like STTs in the right electrode of FGT|☐|FGT AFMTJ as a function of bias-voltage.

Next, we consider a different type of AFM metal, namely the recently discovered two-dimensional (2D) van der Waals magnet Fe$_4$GeTe$_2$, where the AFM order is introduced by doping [76]. A monolayer Fe$_4$GeTe$_2$ contains seven atomic layers stacked with a Te-Fe-Fe-Ge-Fe-Fe-Te sequence and Fe moments coupled ferromagnetically (Fig. 3(a)). A bilayer Fe$_4$GeTe$_2$ has two AFM ordered monolayers. It has $\hat{P}\hat{T}$ symmetry and hence Kramers degeneracy of its band structure (Fig. 3(b)). Such antiferromagnets are considered to host a hidden spin polarization [77]. The bilayer Fe$_4$GeTe$_2$ has a large $d_{AB}/d_{AA} > 2.82$, indicating a very weak inter-sublattice coupling. For the transport along the in-plane $[1\bar{1}0]$ direction, we predict a Néel spin current with a large spin polarization $p_{\alpha\parallel} = 68\%$.

In order to illustrate possible spin-dependent transport phenomena driven by the Néel spin currents in such spin-degenerate antiferromagnets, we build two artificial AFMTJs (denoted as FGT|GeTe|FGT and FGT|☐|FGT) with Fe$_4$GeTe$_2$ $(1\bar{1}0)$ electrodes, and use GeTe molecules as the barrier for FGT|GeTe|FGT (Fig. 3(c)) and a vacuum layer (denoted by ☐) of ~4 Å for FGT|☐|FGT (Fig. 3(d)). Figure 3(e,f) shows the calculated transmissions of these AFMTJs as functions of energy for parallel (P) and antiparallel (AP) states of the Néel vector. We find that independent of the energy and the barrier, the total transmission is always larger for the P state ($T_P$) than for the AP state ($T_{AP}$) in the calculated energy window. At the Fermi energy ($E_F$), the TMR ratio $(T_P - T_{AP})/T_{AP}$ is ~93% for FGT|GeTe|FGT AFMTJ and ~24% for FGT|☐|FGT AFMTJ (Fig. 3(g)). These sizable TMR values are unexpected since the Fermi surfaces of the electrodes are spin degenerate for the P and AP states. On the other hand, it can be qualitatively understood by Julliere's formula $TMR = \frac{2p_{\alpha\parallel}^2}{1-p_{\alpha\parallel}^2}$ [3, 4] with the large $p_{\alpha\parallel}$, indicating that this spin neutral AFMTJ can be qualitatively considered as two MTJs connected in parallel. Figure 3(h) shows the calculated STT for this AFMTJ. We find a large $\tau_{FL}$ and negligible $\tau_{DL}$, due to the staggered spin polarizations in the Néel spin currents.

Finally, we show that the Néel spin currents are crucial for the deterministic switching of the Néel vector in antiferromagnets. We perform a macroscopic spin dynamics simulation in an antiferromagnet with a $z$-directional anisotropy field $H_K = 0.1$ T, an exchange field $H_{ex} = 100$ T, and a damping parameter $\mu = 0.01$. A current is applied to generate an effective field $H_\alpha$ in sublattice $\alpha$ and hence $\tau_{FL,\alpha}$ and $\tau_{DL,\alpha}$ for switching. If there are only perfect staggered Néel spin currents, $H_\alpha$ is staggered as $H_A = -H_B = H_N$. A uniform spin current contributed by the interface generates an additional effective field $H_U$ and influences $H_\alpha$ as $H_A = H_N + H_U$ and $H_B = -H_N + H_U$. In this case, $\tau_{FL,\alpha}$ and $\tau_{DL,\alpha}$ can be decomposed to uniform $\tau_{FL\_N,\alpha}$ and staggered $\tau_{DL\_N,\alpha}$ driven by the Néel spin currents, and staggered $\tau_{FL\_U,\alpha}$ and uniform $\tau_{DL\_U,\alpha}$ driven by the uniform spin currents.

We fix $H_N = 0.2$ T and change the magnitude of $H_U$ in the simulation. In the absence of $H_U$, we find that the Néel vector initially pointing along the $+z$ direction is rapidly reversed in a few ps (Fig. 4(a)), indicating an ultrafast deterministic switching induced by the Néel spin currents. We also perform the simulation solely with $\tau_{FL\_N,\alpha}$ or $\tau_{DL\_N,\alpha}$, and confirm that the uniform $\tau_{FL\_N,\alpha}$ is responsible for the switching [66]. For a small $H_U = 0.05$ T, $\tau_{DL\_U,\alpha}$ induces an oscillation of the Néel vector which gradually decays during the current application (Fig. 4(b)). This leads to a longer time for the full switching of the Néel vector. For a larger $H_U = 0.1$ T, we find that the oscillation is persistent and hence the full switching of the Néel vector cannot be realized by a constant current (Fig. 4(c)), indicating the dominant role of $\tau_{DL\_U,\alpha}$ in this case. We note, however, that the $z$-component of the Néel vector keeps being negative during the oscillation because of $\tau_{FL\_N,\alpha}$. Therefore, if we apply a short current pulse (10 ps as shown in Fig. 4(c)), the Néel vector will first oscillate and then rapidly relax to the $-z$ direction after the pulse ended. These simulations clearly prove the decisive role of the Néel currents in the deterministic Néel vector switching.

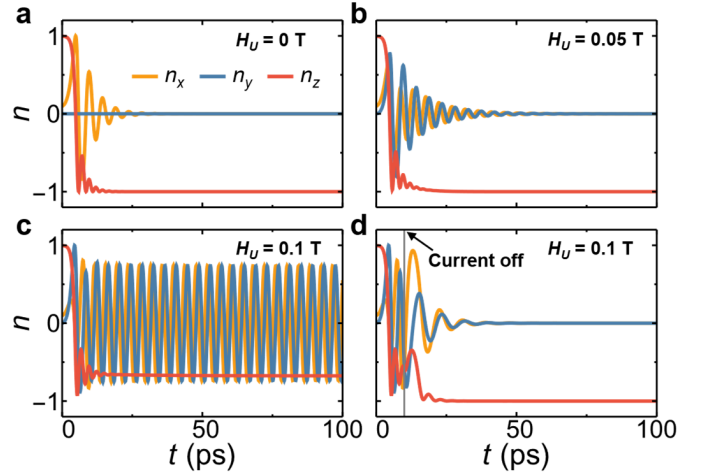

**FIG. 4:** The simulated dynamics of the $x$-, $y$-, and $z$-components of the Néel vector under the current induced effective fields $H_A = H_N + H_U$ and $H_B = -H_N + H_U$. **(a)** $H_N = 0.2$ T and $H_U = 0$ T. **(b)** $H_N = 0.2$ T and $H_U = 0.05$ T. **(c)** $H_N = 0.2$ T and $H_U = 0.1$ T. **(d)** $H_N = 0.2$ T and $H_U = 0.1$ T for the first 10 ps and then $H_N = H_U = 0$.

The STT and TMR driven by Néel spin currents are feasible in experiment. Especially, epitaxial films of AFM RuO$_2$ have been grown experimentally [29, 40-42], promising for the high-quality RuO$_2$-based AFMTJs. The recently developed edge-epitaxy techniques allow fabricating AFMTJs based on lateral heterostructures using the A-type 2D AFM metals such as Fe$_4$GeTe$_2$ [78-80]. Moreover, the Néel spin currents are also expected to generate self-torque in these materials in the presence of the asymmetric boundary conditions. This is analogous to the self-torque driven by a uniform spin current in a single ferromagnet [81,82], and is particularly useful for the AFM spintronics based on a single antiferromagnet [57].

The factor $d_{AB}/d_{AA}$ can serve as a simple measure of the Néel spin current strength in other antiferromagnets different from RuO$_2$ and Fe$_4$GeTe$_2$. We note that some antiferromagnets with a large $d_{AB}/d_{AA}$ might be insulating. While, in this case, generating the Néel spin currents by an electric field is difficult, it is possible to use heat or laser light to generate spin-neutral thermoelectric or photogalvanic currents that support the Néel spin currents. This opens new opportunities for spin caloritronics [83] and optospintronics [84] based on insulating antiferromagnets. The Néel spin currents can be also promising in conventional spintronics by separating the Néel spin currents of two sublattices. For example, attaching an AFM bilayer Fe$_4$GeTe$_2$ to the top of a nonmagnetic material can serve as a contact of a spin transistor, where only the Néel spin current at the bottom layer is injected into the transistor.

In conclusion, we have proposed that the spin-neutral currents in antiferromagnets can host the staggered Néel spin currents flowing through different magnetic sublattices. Based on first-principles quantum-transport calculations, we have demonstrated that RuO$_2$, an antiferromagnet with the spin-split Fermi surface, and Fe$_4$GeTe$_2$, an antiferromagnet with the spin-degenerate Fermi surface, support strong Néel spin currents that generate a giant TMR and sizable field-like STT, the latter being crucial for the deterministic switching of the Néel vector. Our work uncovers a previously unexplored potential of antiferromagnets, and paves a new route to realize efficient reading and writing of the Néel vector for antiferromagnetic spintronics.

**Acknowledgments.** We thank Qihang Liu and Bo Li for helpful discussions. This work was supported by the National Key Research and Development Program of China (Grant Nos. 2021YFA1600200, 2022YFA1403203), the National Science Foundation of China (NSFC Grants No. 12274411, 12174019, 12274412, 12274115, 12204009, U2032215), and the Program for Science & Technology Innovation Talents in Universities of Henan Province (No. 23HASTIT027). The calculations were performed at Hefei Advanced Computing Center. The figures were created using the SciDraw scientific figure preparation system [85].

* These authors contributed equally to this work.

† dfshao@issp.ac.cn

‡ tsymbal@unl.edu